\documentstyle[emulateapj]{article}

\begin{document}
\submitted{Accepted by The Astrophysical Journal Letters, 1999 July 12}
 
\title{An Upper Limit on the Reflected Light\\
	from the Planet Orbiting the Star $\tau$~Bootis}
 
\author{David Charbonneau, Robert W. Noyes, Sylvain G. Korzennik,\\ 
Peter Nisenson and Saurabh Jha}
\affil{Harvard-Smithsonian Center for Astrophysics, 
60 Garden Street, Cambridge, MA 02138; dcharbonneau@cfa.harvard.edu}
\and
\author{Steven S. Vogt and Robert I. Kibrick}
\affil{UCO/Lick Observatory, University of California at Santa Cruz,
Santa Cruz, CA 95064}
 
\begin{abstract}
The planet orbiting $\tau$~Boo at a separation of 0.046 AU 
could produce a reflected light flux as bright
as $1 \times 10^{-4}$ relative to that of the star.  A spectrum of
the system will contain a reflected light component which
varies in amplitude and Doppler-shift as the planet orbits the
star.  Assuming the secondary spectrum is primarily the 
reflected stellar spectrum, we can limit the relative reflected light
flux to be less than $5 \times 10^{-5}$.
This implies an upper limit of 0.3 for the planetary geometric
albedo near 480~nm, assuming a planetary radius
of 1.2 $R_{\rm Jup}$.  This albedo is significantly less
than that of any of the giant planets of the solar system,
and is not consistent with certain published theoretical predictions.
\end{abstract}


\section{INTRODUCTION}

Radial velocity surveys of nearby F, G, K and M dwarf stars have
revealed eight planets (Mayor \& Queloz 1995; Butler et al. 1997; 
Butler et al. 1998; Fischer et al. 1999; Mayor et al. 1999;
Queloz et al. 1999) which orbit their parent stars with a separation
of $a \lesssim 0.1 \, \rm{AU}$.  These close-in extrasolar giant
planets (CEGPs) may be directly detectable by their reflected light,
due to the proximity of the planet to the illuminating star.  In this Letter,
we present the results of a spectroscopic search for the reflected light component 
from the planet orbiting the star $\tau$~Boo.  The motivation to attempt
such a detection for a CEGP is strong:  It would constitute the first 
direct detection of a planet orbiting another star.  It would yield the
orbital inclination, and hence the planetary mass, and would also
measure a combination of the planetary radius and albedo, from
which a minimum radius can be deduced.  Furthermore, it would open the
way to direct investigation of the spectrum of the planet itself.  
Conversely, a low enough upper limit
would provide useful constraints on the radius and albedo of the CEGP.

\section{REFLECTED LIGHT}

\subsection{Photometric Variations}
In order to calculate the predicted flux ratio of the planet
relative to the star, let $R_p$ denote
the planetary radius, $R_s$ the stellar radius, $a$ the physical
separation, and $\alpha$ the angle between the star
and the observer as seen from the planet.  The observationally
useful quantity is the geometric albedo $p$, 
which is the flux from the planet at $\alpha = 0$
divided by the flux that would be measured from a Lambert law 
(ie. perfectly diffusing; see, for example, Sobolev 1975) disk of the 
same diameter, located at the distance of the planet.  
In the case that ${R_p} \ll {R_s} \ll {a}$, 
the ratio $\epsilon$ of the observed flux from the planet at $\alpha = 0$
to that of the star is
\begin{equation}
\epsilon = p \, \left( \frac{R_p}{a} \right)^2.
\label{eps}
\end{equation}
The value of $p$ depends on the amplitude and angular
dependence of the various sources of scattering in the
planetary atmosphere, integrated over the surface of the sphere.
For a Lambert law sphere, $p = 2/3$, whereas for a semi-infinite 
purely Rayleigh scattering atmosphere, $p = 3/4$.  
The geometric albedos at 480~nm of Jupiter, Saturn, Uranus and Neptune are 
0.46, 0.39, 0.60, and 0.58, respectively (Karkoschka 1994).

We treat the orbit as circular, since the 
observed orbit of $\tau$~Boo has an 
eccentricity less than 0.02 (Butler et al. 1997).  
We neglect occultations, since a transit
would produce a $\sim 0.01 \, {\rm mag}$ 
photometric dip, and is ruled out 
by Baliunas et al. (1997).  We take the orbital phase $\Phi \in [0,1]$ to
be 0 at the time of maximum radial velocity of the star.  The phase angle 
$\alpha \in [0,\pi]$ is then defined by
\begin{equation}
\cos {\alpha} = - \sin {i} \sin {2 \pi \Phi}
\label{alp}
\end{equation}
where $i \in [0,\pi/2]$ is the orbital inclination.
The flux from the planet at a phase angle $\alpha$ relative to that
at $\alpha = 0$ is denoted by the phase function $\phi(\alpha)$.
In the case of a Lambert law sphere, the phase-dependent flux ratio $f(\Phi,i)$ 
is given by (Sobolev 1975)
\begin{equation}
f(\Phi, i) = \epsilon \, \phi(\alpha) =  p \, \left( \frac{R_p}{a} \right)^2 \, 
\biggl[ \frac{\sin {\alpha} + (\pi - {\alpha})
\cos {\alpha}}{\pi} \biggr].
\label{f}
\end{equation}
For this analysis, we assume the phase variation of the reflected light
is described by equation \ref{f}.
The phase functions of the gas giants of our solar system 
are well approximated as Lambert spheres (see, for example, Pollack et al. 1986). 

In the case of $\tau$~Boo, Baliunas et al. (1997) can exclude a 
sinusoidal photometric variation at the planetary orbital period with a
peak-to-peak amplitude of 0.4 millimag or greater.  
The predicted variation due to a highly reflective companion
of Jupiter size is $\sim$ 0.1 millimag.
If proposed photometric satellite missions (Matthews 1997; Rouan et al. 1997) 
can achieve a precision of $\sim 10 \, \mu{\rm mag}$ with stability over 
timescales of a few days, 
they could measure this photometric modulation, as discussed
by Charbonneau (1999a).  

\subsection{Spectroscopic Variations}
We assume that $\tau$ Boo has a stellar mass of $M_{s} = 1.2 M_{\sun}$, based
on its spectral classification as an F7~V star (Perrin et 
al. 1977).  It has a ($B-V$) color of 0.48, consistent with the
spectral classification.  The radial velocity 
observations (Butler et al. 1997; Marcy 1997) 
provide the orbital period ($P = 3.3125 \, {\rm d}$),
phase ($T_{\Phi=0} = 2450526.916 \, {\rm JD}$), eccentricity ($e = 0$) 
and amplitude ($K_{s} = 468 \, {\rm m \ s^{-1}}$), from which a semi-major axis
of $a = 0.0462 \, {\rm AU}$ and a planetary mass of 
$M_{p} = 3.89 M_{\rm Jup} / \sin {i}$ are calculated.
The radial velocity of the planet relative to that of the star is 
\begin{equation}
v_p(\Phi, i) = - K_s \, \frac{M_s + M_p}{M_p} \, \cos {2 \pi \Phi}.
\label{vp}
\end{equation}
This has a maximum amplitude of $| v_p(\Phi, i) | \simeq 152 \ {\rm km \, s^{-1}}$.

Thus the spectrum of the system could contain a secondary component which
varies in amplitude according to equation \ref{f} and in Doppler-shift according to
equation \ref{vp}.  Charbonneau, Jha \& Noyes (1998) demonstrate that the effect of
the reflected light component on the line profile bisector is not far from
current observational limits for a CEGP of high reflectivity.  This is 
an alternate technique which may be used to directly detect or limit
the reflected light from a CEGP.

\subsection{Tidal Locking Effects}
Baliunas et al. (1997) use the activity-rotation relation of Noyes et al. (1984) 
and the mean Ca {\small II} flux of $\tau$ Boo to predict a stellar rotation period of 5.1 days.
Analysis of the observations of the Ca {\small II} H \& K lines by 
Baliunas et. al (1997) 
yield a weakly detected period of 3.3 $\pm$ 0.5 d, consistent with the observed 
orbital period of the 
planet, implying that the star and planet form a tidally locked system.
Marcy et al. (1997) demonstrate that, in the case of $\tau$~Boo, a convective envelope 
of mass $M_{CE} \approx 0.01 M_{\sun}$ could become tidally locked in less than the
age of the system.  If so, then there is no relative motion of any point
on the surface of the planet relative to any point on the surface of the star.
In this case, the planet reflects a non-rotationally-broadened stellar spectrum, with
a typical line width dominated by the stellar photospheric convective motions 
($\sim 4 \ {\rm km \, s^{-1}}$; Baliunas et al. 1997).
Thus, one might expect relatively narrow planetary lines superimposed on
much broader stellar lines.

\section{TARGET SELECTION AND OBSERVATIONS}

Several considerations entered into the choice of $\tau$~Boo (HR~5185, HD~120136)
as the optimal candidate for this experiment.  Firstly, the semi-major axis of 
$\tau$ Boo was smaller than that of the other three CEGPs (51 Peg, $\upsilon$ And, \& 
$\rho^{1}$ Cnc) known at the time, which is desirable since the relative amplitude of 
the reflected light decreases with the square of the planet-star distance.  
Secondly, the visual brightness
of $\tau$~Boo is greater than either 51 Peg or ${\rho}^{1}$ Cnc.
The photon noise of the star is the dominant source of noise
in the experiment, and a brighter star allows for a more
precise determination of the stellar flux in a given amount
of observing time.  Thirdly, as discussed above, it may be that
the star is rotating with the planetary orbital period.  
If so, the planetary spectral features would be much sharper and deeper than
those of the primary, which might facilitate their separation.

We observed $\tau$ Boo for three nights (1997 March 20 to March 22) 
using the HIRES echelle spectrograph (Vogt et al. 1994) mounted on the 
Keck-1 Ten-Meter Telescope at the W.~M.~Keck~Observatory located 
atop Mauna~Kea in Hawaii.  These nights were carefully
selected based on the phase of the companion's orbit.  
The spectral range used in this analysis was 465.8~nm to 498.7~nm, and the observations
were made at a resolution $R \equiv \lambda / \delta \lambda$ of either 
60~000 (March 20) or 45~000 (March 21 \& 22).

Since the apparent magnitude of $\tau$~Boo is 4.5 mag, the high flux from the
star would saturate the detector pixels for an exposure time less than
the readout time of the CCD.  To avoid this readout-time-limited scenario,
the cross-disperser was slowly trailed during each observation 
so as to spread the photons over roughly 30 pixels.  This allowed for typical
exposure times of 300 seconds, which resulted in a duty cycle of $\sim 70\%$.  
In all, 154 spectra of $\tau$ Boo were obtained, with a nightly breakdown of
38 for March 20, 32 for March 21 and 84 for March 22.  Cloudy weather 
degraded the number and quality of the spectra on March 21.

\section{DATA ANALYSIS}

\subsection{Extraction of the One-Dimensional Spectra}
Since the extraction of the one-dimensional spectra from the two-dimensional 
exposures must be accomplished without introducing systematic errors above
the level of $1 \times 10^{-4}$ per dispersion element, it was necessary to create
an entirely new and independent set of extraction codes specific to this
experiment.  By so doing, we were able to treat the sources of systematic noise 
particular to the Keck HIRES and these observations, as well as have the
necessary control in identifying sources of contamination as our understanding
of the data proceeded.  

To extract the one-dimensional spectra from an individual frame,
the following algorithm was applied:  The bias was subtracted and the 
non-linear gain was corrected.  A two-dimensional scattered light model was
derived by fitting the inter-order scattered light, and subtracted.  The
two-dimensional flat-field correction was applied, and the orders were extracted by
summing along the cross-dispersion direction, making use of windows which
we had produced to identify the location of both the spectral orders and
the cosmetic defects from internal reflections and a felt-tip pen mark.  
The one-dimensional spectra were then corrected for cosmic rays by cubic 
spline interpolation across contaminated pixels.  A low amplitude 
source of high frequency noise in the extracted spectra 
was corrected for by applying a narrow notch filter in Fourier space.
The typical signal-to-noise-ratio (SNR) was $\sim 1500$ per dispersion element.
The wavelength solution was derived from extracted Th \& Ar emission line
spectra taken throughout the observing run.

\subsection{The Model}
The model is that the data contain a secondary spectrum, 
identical to that of the primary, but Doppler-shifted
due to the orbital motion of the planet and varying in amplitude with 
the angle subtended between the star, the planet and the observer.
The key to the method is to first produce a stellar template spectrum, 
and then make use of the orbital parameters from the radial velocity
observations to calculate a model for a given observation taken at a particular
orbital phase.  The methods we briefly describe here will be presented
in detail in an upcoming paper (Charbonneau 1999b).

The high SNR stellar template spectrum was produced for
each of the two instrumental resolutions by combining all of the extracted spectra.
Initially, a high SNR spectrum
was chosen and an optimized model was found which corrected each observation
to this reference (allowing for variations in the wavelength solution and 
instrumental profile, and low-frequency spatial variations of the continuum).  
A summed stellar template spectrum was produced, 
and this process was iterated twice, beyond which point the errors 
were no longer significantly reduced by further iteration.
The errors in the summed stellar template were 1.2 times the
expectation from photon noise, indicating a precision
of $\sim 1 \times 10^{-4}$ per dispersion element.
We note that this may well comprise
the most precise visible stellar spectrum for any star other than the Sun.

For each observed spectrum, 
we first modify the stellar template spectrum in order to correct for 
the aforementioned variations
in the wavelength solution and instrumental profile, and low-frequency
spatial variations of the continuum.
Note that we wish to interpolate the stellar template spectrum, and not perform
the reverse procedure and interpolate the observed spectra, since the
stellar template spectrum is at a much higher SNR.
Then, if we denote by $S$ this modified stellar template spectrum and by $\lambda$ the
wavelength solution, the model at a given pixel $j$ of an observed spectrum
taken at an orbital phase $\Phi$ is described by
\begin{equation}
M_{j} = \frac{S\left(\lambda_{j}\right) + \epsilon \ \phi(\Phi, i) \ S\left(\lambda_{j}
\left[1+\frac{v_p(\Phi,i)}{c}\right]\right)}{1 + \epsilon \ \phi(\Phi, i)}
\label{themodel}
\end{equation}
The two unknown parameters are $\{ \epsilon, i\}$.
The situation in which there in no reflected light from the planet
is equivalent to $\epsilon = 0$.  In this case, the observed spectra are
best fit as replicated stellar spectra.  

As noted earlier, the stellar rotation period may be 
the orbital period of the planet, and hence the reflected spectrum
may be composed of non-rotationally-broadened lines.
The instrumental resolution will smear all spectral lines to a width
of $\sim 7 \ {\rm km \, s^{-1}}$.  Several exposures of the
sharp-lined F8~V star 36~UMa (HR~4112, HD~90839, $V=4.84$, $B-V=0.52$)
were combined and corrected to the Doppler-shift of $\tau$ Boo 
to produce a stellar template spectrum, $S'$.  The spectral differences between an
F7~V and an F8~V star are insignificant for the purposes of this analysis.
The spectrum of 36~UMa serves 
as an excellent mock-up for the non-rotationally-broadened spectrum of $\tau$ Boo 
and includes the instrumental effects.  
Thus we also investigated the model
\begin{equation}
M'_{j} = \frac{S\left(\lambda_{j}\right) + \epsilon \ \phi(\Phi, i) \ 
\gamma \ S'\left(\lambda_{j}
\left[1+\frac{v_p(\Phi,i)}{c}\right]\right)}{1 + \epsilon \ \phi(\Phi, i)}
\label{themodel2}
\end{equation}
where $\gamma$ is a normalization factor.

The model was evaluated by calculating the ${\chi}^2$ parameter 
as a function of $\{ \epsilon, i \}$.
The minimum ${\chi}_{\rm min}^2$ is subtracted off to define
$\Delta {\chi}^2 = {\chi}^2 - {\chi}_{\rm min}^2$.
The confidence levels in the allowed values of the parameters 
are described by drawing contours of fixed $\Delta {\chi}^2$
at a desired set of significance levels.
The confidence levels were tested for a given choice of $\{\epsilon,i\}$ 
by directly injecting a reflected light secondary at the correct amplitude and Doppler-shift
into each observed spectrum.  At $\epsilon \gtrsim 10^{-3}$ and 
high inclination, the secondary can be detected at the 99\% confidence level
with only one spectrum.  At $\epsilon \simeq 10^{-4}$,
the planet is recovered only by considering all of the spectra, and 
the uncertainty in the parameters is significantly greater.  
Tests showed that the planet could be recovered for $i \gtrsim 10^{\circ}$.  

A second test was provided by the detection of solar contamination
employing a model similar to the one described in equation \ref{themodel2}, but with
the modification that the secondary spectrum is at a constant
(but unknown) amplitude and Doppler-shift.  Solar contamination
was detected at the Doppler-shift between the Sun and $\tau$ Boo, and at a 
relative amplitude of $10^{-3}$, in the spectra taken on March 21.
The source of this contamination was reflection of the solar spectrum
off the Moon and subsequently off the clouds which were present throughout the
night.  The exclusion of the contaminated spectra from the reflected light analysis 
did not greatly reduce the statistical significance 
as these spectra contained only 10\% of the photons of the entire data set.

\section{RESULTS AND DISCUSSION}

We find no evidence for a highly reflective planet orbiting $\tau$~Boo.  
For $i \gtrsim 10^{\circ}$,
we can constrain the reflected flux ratio $\epsilon \lesssim 8 \times 10^{-5}$ 
at the 99\% confidence level, under the assumptions that the 
reflected light spectrum is a copy of the stellar spectrum.  
For $i \gtrsim 70^{\circ}$, this improves to $\epsilon \lesssim 5 \times 10^{-5}$.  
Assuming a planetary radius of 1.2 $R_{\rm Jup}$ (Guillot et al. 1996), this limits 
the geometric albedo to $p \lesssim 0.3$. 
Figure~1 shows the precise limit of the reflected light amplitude
as a function of orbital inclination.  Under the
assumption that the secondary reflects a non-rotationally-broadened version
of the stellar spectrum, this limit becomes stronger for high inclinations.
The particular shape of a given confidence level in Figure~1 results from the interplay 
of the orbital phases and statistical weights of the set of spectra.  
The upper limit imposed is set primarily 
by the last night of observing (March 22), when the planet was near a phase
of $\Phi = 0$.  The dip down to stronger constraints on the flux
ratio at an inclination $i \simeq 15^{\circ}$ results 
from the first night of observing (March 20) when the planet was near inferior conjunction:
Only if the planet is at low inclinations will it be expected to contribute
a reasonable reflected light signal and hence allow us to significantly differentiate
between models.  

At very low inclinations ($i \lesssim 10^{\circ}$), this experiment is not able to
exclude even very bright companions due to both the lack of a significant 
Doppler-shift between the primary and the secondary, and the lack of a 
phase variation in the light from the secondary.  However, 
these low inclination orbits may be excluded under a further consideration:  If
the axis defined by the stellar rotation is the same as that of the
orbit of the planet, then the observed $v \, \sin {i} \simeq 15 \ {\rm 
km \, s^{-1}}$ for the star would imply a true rotational velocity of greater
than 50 $\rm{km \, s^{-1}}$ for $i \lesssim 17^{\circ}$.  Such high rotational velocities
are not observed (Gray 1982) for main-sequence F7 stars.  
High inclination orbits can be excluded by the lack of eclipses from 
photometric monitoring.  Baliunas et al. (1997) exclude $i \gtrsim 83^{\circ}$.  
This is consistent with our experiment as we find no evidence for a companion at these
high inclinations.

We reiterate that the derivation of an upper limit for the geometric albedo
requires the assumption of a value for the planetary radius (1.2 $R_{p}$) 
and a functional form for the phase variation (a Lambert law sphere).
If the actual values are significantly different than these, then the upper
limit on the geometric albedo is modified as well.  For example, assuming 
a smaller planetary radius would permit a larger albedo (see equation \ref{eps}).

Published predictions of the albedo of CEGPs vary by many orders of magnitude,
and are highly sensitive to the presence of condensates in the planetary
atmosphere.  Burrows \& Sharp (1999) consider cloud
formation and depletion by rainout, and 
demonstrate that $\rm{MgSiO_{3}}$ will be an abundant condensate
at the effective temperature of $\tau$ Boo b ($\sim 1500 \, \rm{K}$).  Marley et al. (1999)
calculate both cloud-free and silicate cloud atmospheres and 
predict $0.35 \lesssim p(480 \, {\rm nm}) \lesssim 0.55$ for an EGP with a temperature
of 1000 K, which is greater than our upper limit of $p(480 \, {\rm nm}) = 0.3$.  
They neglect the effects of stellar insolation on the model
atmosphere.  Seager \& Sasselov (1998) explicitly include the stellar flux, 
solve the equation of 
radiative transfer through a model atmosphere of $\tau$ Boo b, and predict
$p(480 \, {\rm nm}) \simeq 0.0002$.  The low albedo is due in part to the absorption of
photons by TiO in the atmosphere.  However, it may be that the TiO forms and rains out, 
and thus is not an important factor.  Including the presence of $\rm{MgSiO_{3}}$ clouds,
Seager \& Sasselov predict a larger (but still very dark) albedo of $p(480 \, {\rm nm}) \simeq 
0.003$.  The reflectivity of the $\rm{MgSiO_{3}}$ grains at a given wavelength
is highly dependent on the grain size relative to the wavelength of light.
Burrows \& Sharp (1999) also predict that other condensates (such as Fe) may be present
at these temperatures.  If iron droplets are a significant condensate, the resulting albedo
would be very dark due to the high absorption at optical wavelengths.  
Given the current uncertainty in the models, there are many reasonable model planetary 
atmospheres which are consistent with our upper limit.

We have achieved the current upper limit using only a limited
spectral range, and data obtained when the planet was far from
opposition.  It is restricted by the photon noise of the
data set, not by systematic errors.  By expanding the spectral range
and observing on several nights when the planet is near opposition,
it would be possible to significantly reduce this upper limit.
It may be advantageous to conduct this experiment at shorter wavelengths, 
since Seager \& Sasselov (1998) predict a dramatic rise in the albedo
shortwards of 420 nm.  

\acknowledgements
We gratefully acknowledge the NASA/Keck Time Assignment Committee 
for the observing time allocation.
This work was supported in part by NASA Grant NAG5-75005.
%\newpage

\newpage

\begin{figure}
\plotone{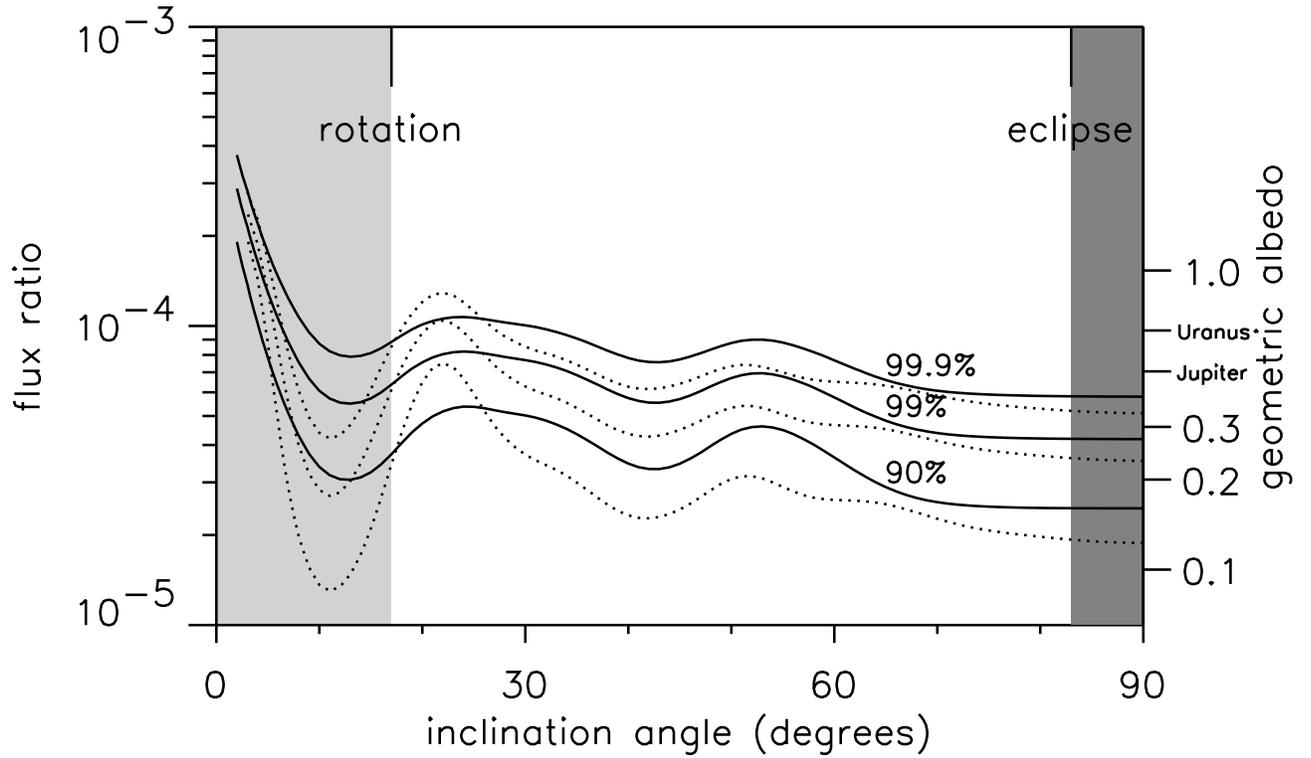}
\figcaption{The solid curves are the 90\%, 99\%, \& 99.9\% confidence levels
on the upper limit for the relative reflected flux $\epsilon$ as
a function of orbital inclination $i$, if the 
reflected light is a copy of the stellar spectrum. 
The dashed curves are the same confidence levels under the
assumption that the system is tidally locked and thus
the planet reflects a non-rotationally-broadened copy of the stellar spectrum.  
Upper limits on the geometric albedo $p$ under the assumption that
$R_{p} = 1.2 \, R_{\rm Jup}$ are shown on the right-hand axis,
and the values for Jupiter and Uranus are included for comparison. 
The lack of transits excludes $i \gtrsim 83^{\circ}$, and $i \lesssim 17^{\circ}$
can be excluded under the assumption that the stellar rotation axis is 
co-aligned with that of orbital motion, as discussed in the text.}
\end{figure}

\end{document}